# Enhancement of ferroelectricity in Cr-doped $Ho_2Ti_2O_7$

X. W. Dong[1], S. Dong[2], K. F. Wang[1], J. G. Wan[1], and J. –M. Liu[1,3 &]

[1]*Laboratory of Solid State Microstructures, Nanjing University, Nanjing 210093, China*
[2]*Department of Physics, Southeast University, Nanjing 211189, China*
[3]*International Center for Materials Physics, Chinese Academy of Sciences, Shenyang, China*

[Abstract] A series of polycrystalline pyrochlore rare-earth titanate $Ho_{2-x}Cr_xTi_2O_7$ are synthesized in order to enhance the ferroelectricity of pyrochlore $Ho_2Ti_2O_7$. For the sample close to the doping level $x=0.4$, a giant enhancement of polarization $P$ up to $660\mu C/m^2$ from $0.54\mu C/m^2$ at $x=0$ is obtained, accompanied with an increment of ferroelectric transition point $T_c$ up to ~140K from ~60K. A magnetic anomaly at $T$~140K together with the polarization response to magnetic field, is identified, implying the multiferroic effect in $Ho_{2-x}Cr_xTi_2O_7$.

---

[&] Author to whom all correspondences should be addressed, E-mail: liujm@nju.edu.cn

Multiferroics, in which ferroelectric (FE) and spin orders coexist and couple with each other through magnetoelectric (ME) effect, have recently become a focus of many researches due to the potential applications as well as challenges to fundamental physics.[1-4] Up to date, more and more attentions have been paid to search for novel multiferroics with high FE Curie point ($T_c$), large FE polarization ($P$), and intrinsic ME coupling.[5-10] Recently, we reported the coexisting spin and FE behaviors in pyrochlore compound $Ho_2Ti_2O_7$ (HTO) below $T_c$~60K, although the measured $P$ below temperature $T$~10K is very weak (~0.54 $\mu C/m^2$).[11]

HTO is a member of pyrochlore materials with the general formula $A_2B_2O_7$, and crystallizes in pyrochlore structure (space group $Fd$-$3m$).[12] The ideal pyrochlore crystal structure can be considered as two independent three-dimensional frameworks ($A_2O'$ and $B_2O_6$), thus $A_2B_2O_7$ is often written as $A_2B_2O_6O'$ to distinguish the oxygen atoms of two different networks.[13] As for the details, the $B_2O_6$ framework consists of $[BO_6]$ octahedral sharing all vertices to form large cavities, and interpenetrates with $A_2O'$ framework such that the O' ions of the $A_2O'$ occupy the centers of these cavities, while the A cations reside in the puckered hexagons formed by oxygen atoms in the $B_2O_6$. As a spin-ice compound, $Ho_2O'$ sublattice of HTO (Fig.1(a) shows one view of $Ho_2O'$ sublattice) can be decomposed into two orthogonal sets of chains, which run along the orthogonal [1$\bar{1}$0] and [110] directions, respectively. At low-$T$, $Ho^{3+}$ cations occupy the corner sites of tetrahedra, whose easy axes of spins point to the centers of tetrahedra (the local <111> directions), as shown in Fig.1(b).

As far as we know, a well-confirmed fact is that the A cation must be appreciably larger than the B cation in this structure.[13,14] A distortion of the $[BO_6]$ octahedral will be induced if the radius of A cation is small. It is noted that such lattice distortion is very impacting on the FE behaviors of multiferroics. Therefore, one can dope small cation at the A site to modulate (enhance) the lattice distortion, then to enhance the ferroelectricity. As for HTO, the radius of $Ho^{3+}$ ($r$=0.901Å) is extremely large (noting $r$=0.605Å for $Ti^{4+}$), thus allowing us an opportunity to modify HTO in terms of the ferroelectricity improvement. In order to induce a large lattice distortion, we choose small magnetic ions $Cr^{3+}$($r$=0.615Å) to substitute $Ho^{3+}$ sites in HTO.

Moreover, the magnetic ion $Ho^{3+}$ in HTO carries a large magnetic moment (~10$\mu_B$), such that the long-range magnetic dipolar interaction is as significant as the nearest-neighbor exchange.[15] While the spin moment of $Cr^{3+}$ (~$3\mu_B$) is much smaller than $Ho^{3+}$, such a substitution would weaken the long-range magnetic dipolar interactions, and may allow the nearest-neighbor exchange to be dominant, thus offering a different magnetic behavior in Cr-doped HTO. This is another focus of our work. We will investigate the FE and magnetic properties of $Ho_{2-x}Cr_xTi_2O_7$ (HCTO). At last, we will discuss the origins of ferroelectricity through the polarization measurements.

Polycrystalline HCTO samples were synthesized by standard solid state reaction from the oxide starting materials. The phase purity was checked by X-ray diffraction (XRD) using Cu $K\alpha$ radiation and transmission electron microscopy (TEM). The $T$- and magnetic field ($H$) dependences of magnetization ($M$) were measured using the superconducting quantum interference device (SQUID) magnetometer (Quantum Design, Inc). For measuring $P$, we detected the pyroelectric current ($I$) using an electrometer (Keithley 6514) associated with the PPMS. The poling electric field is $E$=5kV/cm. This technique has been repeatedly demonstrated to be reliable for those multiferroics of negligible trapped charges, while the data may be suspicious if the trapped charges or leakage is high. Fortunately, our HCTO samples are of high quality and we calibrate the pyroelectric current as a function of $T$ at various warming ramp rates, excluding the contribution from de-trapped charges.[10] The ferroelectric hysteresis was also evaluated using an electrometer (Keithley 6514) associated with the PPMS, by a systematic measurement of the pyroelectric current and polar-switching current upon reversal of electric field.

First, we check the crystallinity and structure of the as-prepared samples at room temperature and the XRD θ-2θ spectra data for HCTO at $x$=0, 0.1, 0.15, 0.2, 0.25, 0.3, 0.35, and 0.4 are presented in Fig.1(c). It indicates that the samples all have the single-phase pyrochlore structure and no impurity diffraction peak is identified. After a careful analysis, one can see that the reflection peaks shift toward the high θ-range with increasing $x$ and the structure of Cr-doped HTO becomes stable after a continuous lattice contraction until $x$=0.2, as shown in the inset of Fig.1(c), a reasonable fact considering that Cr ion is much smaller than Ho ion. In addition, we address that the Cr-doping seems to induce a weak shoulder appearing at the immediate right-side of the main peak between 2θ=51° and 52°, as shown in the inset of Fig.1(c), indicating the continuously enhanced lattice distortion with increasing *x*, while details of the lattice symmetry breaking if any remain unclear so far.

We also show the TEM observation to support the XRD results, and one example for HCTO at *x*=0.2 is shown in Fig.1(d). The TEM selected area electro-diffraction patterns observation with the incident beam parallel to the [111] orientation is presented in the inset of Fig.1(d). By a careful examination, on one hand, halos in the pattern of HCTO *x*=0.2 are observable with respect to that of HTO, which means that the crystal structure is distorted by Cr-doping. In addition, the evaluated inter-plane separation *d* of <220>, <422>, <311>, <211> are respectively 3.48, 2.04, 2.952, and 3.98 Å, which are smaller than that of pure HTO (*d*=3.57, 2.06, 3.05 and 4.12 Å for <220>, <422>, <311>, <211> of HTO). It also means that the crystal structure of Cr-doped HTO is shrinking. On the other hand, the samples are of high crystalline quality because the pyrochlore structure is highly tailorable.[13]

Subsequently, we characterize the magnetic and FE behaviors. First, we look at the magnetism of selected samples (*x*=0, and 0.2). It is well known that HTO is spin-ice.[16] Fig 2(a) shows the *M-T* curves measured under *H*=0.01T for both zero-field cooled (ZFC) and field-cooling (FC) conditions. For HTO, no significant difference between the ZFC and FC curves is observed, indicating the spin-ice configuration is quite robust. However, for HCTO *x*=0.2, the ZFC and FC curves show a clear separation at *T*~140K upon decreasing *T*, which is a surprising anomaly. We will address this anomaly later and here first analyze the general magnetic response to varying *T*. If using the Curie-Weiss law to fit the data, one observes a roughly good paramagnetic behavior for HTO even down to *T*~2K, indicating the typical spin-ice behavior, as shown by the dash line in Fig.2(b). Nevertheless, for HCTO *x*=0.2, careful observation reveals not only the aforesaid anomaly, but also a departure of the ZFC curve from the fitting line. This departure indicates that the spin-ice configuration of HCTO *x*=0.2 is not as robust as that of HTO in response to weak magnetic field.

To understand this anomaly, we consult to the spin structure. For HTO, the Ho spins are well characterized by the effective classical Ising spins constrained to the local <111> directions below *T*~200K.[16] As a gradual sequence, the Ho spins will be ordering into the "two-in-two-out" alignment at low-*T*. For HCTO, $Cr^{3+}$ ions replace partial $Ho^{3+}$ ions, resulting in the weakening of the local trigonal crystal field and the long-range magnetic dipolar interactions responsible for the spin ice behavior. This may lead to the appearance of the "three-in–one-out" state, or the local "two-in–two-out" state can be easily changed into the "three-in–one-out" state, driven by e.g. magnetic field.[17-19] Noting that the "three-in–one-out" state has larger magnetization than the "two-in–two-out" state along the [111] direction, one then understands why the *M-T* curves for HCTO slightly deviate from the Curie-Weiss law, due to the partial activation of the "three-in-one-out" state by the measuring *H*. This explains why the spin ice configuration of HCTO is less stable than that of HTO. Moreover, the *M-H* hysteresis loops at *T*~2K for HCTO (*x*=0, 0.2, and 0.4) are shown in Fig 2(c). It is noted that *M* becomes well saturated at *H*=1.0T and the coercive field is very small, a typical feature for spin ice system too. At the same time, Fig 2(d) shows that the measured saturated *M* decreases with increasing *x*, indicating that the Cr-doping suppresses the magnetic moment due to the reason mentioned earlier ($Cr^{3+}$ spin is weaker than $Ho^{3+}$ spin).

Finally, we look at the ferroelectricity. First, the reliability of data is evidenced by the pyroelectric current (*I*) under different warming ramp rates and one example for HCTO *x*=0.2 at the rate of 2K/min and 4K/min is given in Fig.3(a). The measured *I*(*T*) at the two rates are almost coincident with each other, excluding the possible contribution from any trapped charges. Furthermore, Fig.3(b) shows the *T*-dependence of *P* for HCTO at *x*=0.1, 0.15, 0.2, 0.25, 0.3, 0.35, 0.4 with *E*=5kV/cm and also that for HTO as a comparison. For HTO, one can see that the FE transition appears at *T*~60K with an anomalous increment at *T*~23K. The maximum *P* is only ~0.54$\mu$C/$m^2$, quite small with respect to other multiferroics. However, for HCTO at *x*>0.1, the *P* is remarkably enhanced, as also seen in the inset of Fig.3(b) which shows the *x*-dependence of *P* for HCTO samples at *T*~10K. Here, it is noted that *P* at *x*=0.1 is enhanced up to ~1.9$\mu$C/$m^2$ (almost 4 times as that of HTO). But above *x*>0.1, *P* increases hugely with increasing *x* and gets a value as high as 660$\mu$C/$m^2$ at *x*=0.4, which is almost 1200 times as that of HTO. The measured $T_c$ for HCTO *x*=0.4 is ~140K, also much higher than that of HTO.

Moreover, for HCTO *x*=0.2, it is noted that the ferroelectricity generation is coincident with the anomaly of the *M-T* curves, implying the ME effect. This can be further confirmed by the *T*-dependence of Δ*P*=*P*(*H*)-*P*(*H*=0) for HCTO *x*=0.4, measured under *H*=1.0, 3.0, and 9.0T, as shown in Fig.3(c). One can see that Δ*P* increases with increasing *H* over the whole

*T*-range, demonstrating the significant response of *P* against *H*, i.e. the multiferroicity. On the other hand, Δ*P* increases with decreasing *T*, below *T*~140K. This means that the polarization may be modulated by *H*. It is found that Δ*P* is as high as ~42$\mu$C/m$^2$ under *H*=9.0T at low-*T*. Also, the inset of Fig.3(c) shows the measured *P-E* hysteresis for HCTO *x*=0.4 measured at *T*~10K, where electric field *E* is ramping between ±10kV/cm. Although the hysteresis loop is not as good as expected, the feature of HCTO as a FE system is clear.

Based on the above results, we are allowed to clarify the origin of *P* for HCTO. It seems that both the TiO$_6$ octahedra due to the lattice distortion and the Ho spin chains can contribute to the ferroelectricity. We propose the following two possible mechanisms. First, we argue that the shrinking of crystal structure by the Cr-doping induces the distortion of the [TiO$_6$] octahedral and thus the relative shift of Ti$^{4+}$ ions from surrounding oxygen ions, benefiting to the ferroelectricity generation. This mechanism lies in the physics of classical soft-mode phonon and the TO$_4$ mode excitation if any may evidence this picture.[20, 21]

Second, the contribution of inverse Dzyaloshinskii-Moriya interaction (DMI) to the ferroelectricity should also be taken into account.[22] We show the Ho-O-Ho chain of HTO with the Ho spins in Fig.3(d) where the oxygen shift from the Ho chain axis is exaggerated for clarification consideration. It is seen that the local DMI may also contribute to a local electric polar moment. However, due to the Ho-O-Ho chain configuration, the overall polar moment is zero due to the inter-pole cancel along the chain.[23] For HCTO, the continuous Ho-O-Ho chain is broken due to the Cr-doping and the "three-in–one-out" alignment at the Cr-site is induced, generating a spin configuration different from that shown in Fig.3(d). One possible configuration is given in Fig.3(e) where the overall inter-pole cancel becomes unsatisfied, generating the net nonzero polarization.

It should be mentioned that the measured $T_c$ for the FE order is coincident with this magnetic anomaly at *T*~140K, implying that the ferroelectricity generation is indeed related to some extent with the spin configuration too. Unfortunately further direct evidence with the two possible mechanisms responsible for the ferroelectricity generation is needed.

In summary, the polycrystalline pyrochlore Ho$_{2-x}$Cr$_x$Ti$_2$O$_7$ has been prepared and their ferroelectricity and magnetism have been investigated. The magnetic anomaly for HCTO at *T*~140K which is absent for pure HTO has indicated the different spin configuration of HCTO from that of HTO. A remarkably enhanced $P$ upon the Cr-doping is evidenced as $x>0.1$ and the measured $P$ at $x=0.4$ reaches up to $\sim 660 \mu C/m^2$ with $T_c \sim 40K$, which is ~1200 times larger than that of HTO. Also, a weak ME effect below $T\sim 140K$, evidenced by the significant response of $P$ to varying $H$, has been observed. We argue that both the doping-induced lattice distortion and the inverse DMI may benefit to the ferroelectricity in HCTO.

**Acknowledgement**

This work was supported by the National Natural Science Foundation of China (50832002, 10874075), the National Key Projects for Basic Researches of China (2009CB623303, 2006CB929501, 2006CB921802), and the Natural Science Foundation of China in Jiangsu (BK2008024).

**Figure captions**

Fig.1. (color online) (a) Planar view of Ho$_2$O′ sublattice along the [110]. (b) Spin structure of HTO satisfying the ice rules. (c) XRD patterns of HTO and HCTO at $x<0.5$. (d) TEM image of HCTO at $x=0.2$, with the incident beam parallel to [111] and the inset is the selected area electron-diffraction patterns of HCTO ($x=0$ and 0.2) with the incident beams parallel to [111].

Fig.2. (color online) (a) Measured $M(T)$ and (b) $1/M(T)$ dependences under the ZFC and FC sequences for HCTO $x=0$ and 0.2 (measuring field $H=1000$Oe). (c) Measured $M$-$H$ hysteresis for HCTO $x=0$, 0.2, and 0.4 at $T=2$K. (d) Saturated $M$ as a function of $x$ for HCTO.

Fig.3. (color online) (a) Measured pyroelectric current $I$ as a function of $T$ at two warming ramp rates (2K/min, and 4K/min) for HCTO $x=0.2$. (b) Measured $P(T)$ for several HCTO samples and the inset is the $P(x)$ at $T\sim10$K. (c) $\Delta P=P(H)-P(H=0)$ as a function of $T$ for HCTO $x=0.4$ under $H=1.0$, 3.0, and 9.0T, respectively, and the inset shows the $P$-$E$ loop at $T\sim10$K for HCTO $x=0.4$. (d) A sketch of one Ho$_2$O′ chain with the Ho-O-Ho bond angle of 109.5°. (e) A sketch of one possible Ho(Cr)-O-Ho(Cr) chain generating nonzero $P$.

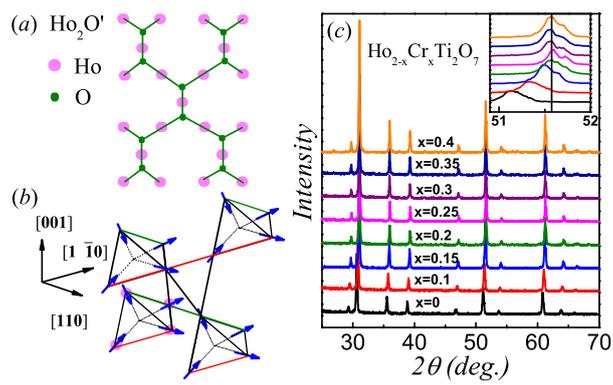

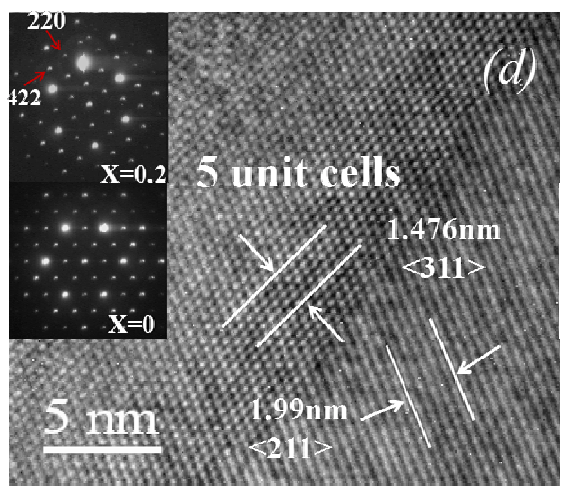

Figure 1

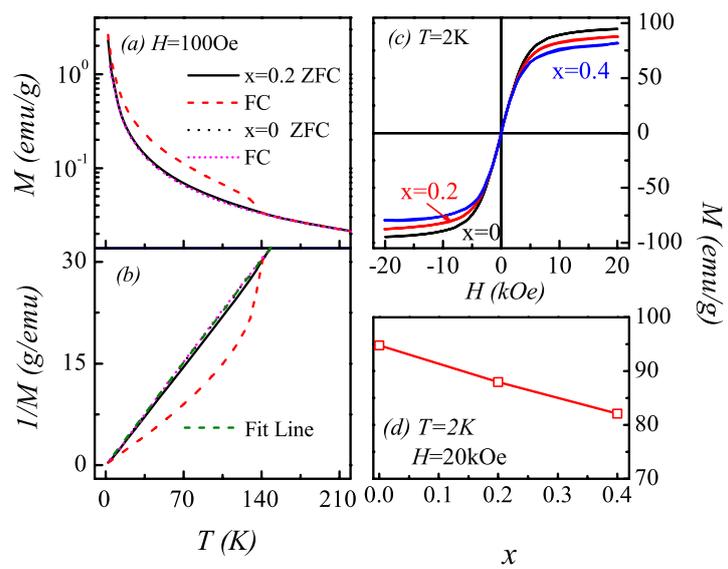

Figure 2

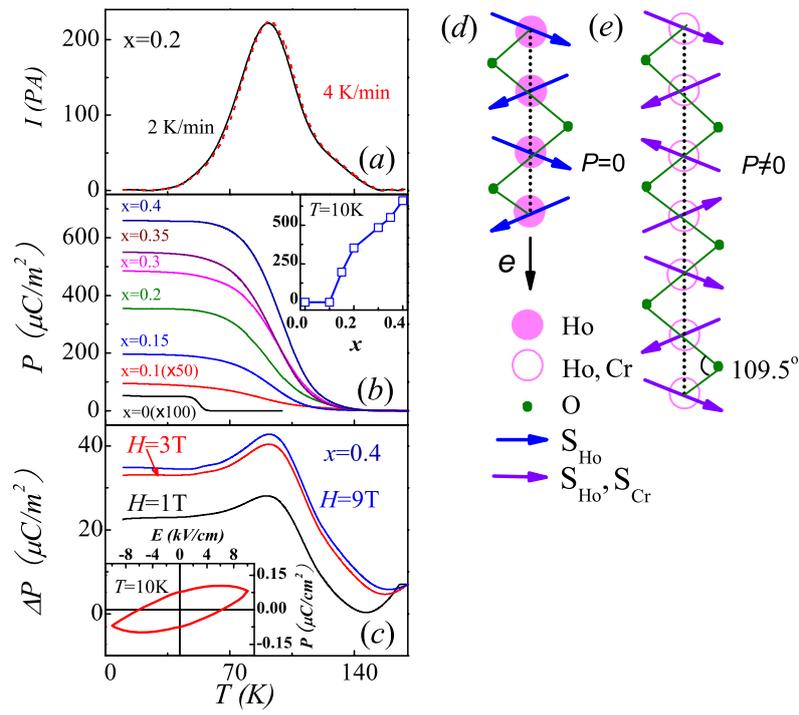

Figure 3